\documentclass{aa}  

\usepackage{graphicx}
\usepackage[dvipsnames]{xcolor}
\usepackage{txfonts}
\usepackage[breaklinks=true]{hyperref} 
\hypersetup{
  colorlinks   = true, 
  urlcolor     = red, 
  linkcolor    = blue, 
  citecolor    = blue, 
  breaklinks   = true 
}
\begin{document}

\title{Detecting gravitational lensing by matter currents }

   \author{C. Murray
          \inst{1,2},
          R. Kou \inst{2,3},
          J. G. Bartlett \inst{2}
          }

\institute{
Université Paris-Saclay, Université Paris Cité, CEA, CNRS, AIM,
91191, Gif-sur-Yvette, France
\and Université Paris Cité, CNRS, Astroparticule et Cosmologie, 75013
Paris, France
\and Department of Physics \& Astronomy, University of Sussex, Brighton BN1 9QH, U.K.
}

   \date{Received September 15, 1996; accepted March 16, 1997}

 
  \abstract
{ We explore the observational prospects for detecting gravitational lensing induced by cosmological matter currents, a relativistic correction to the standard density lensing effect arising from the motion of matter. We propose to isolate this contribution by cross-correlating the weak-lensing convergence field with a reconstructed cosmic momentum field inferred from galaxy redshift surveys. Using numerical simulations, we demonstrate that this reconstructed momentum field is uncorrelated with the density lensing signal, enabling a clean separation of the gravitomagnetic component. We then forecast the detectability of this signal for upcoming wide-field galaxy and weak-lensing surveys, showing that a statistically significant detection may be achievable under realistic observational conditions. Such a measurement would provide the first direct probe of the large-scale cosmic momentum field, offering a novel test of general relativity and Lorentz invariance on cosmological scales. }

   \keywords{ Cosmology: observations --
                Galaxies: clusters
               }

   \maketitle

\section{Introduction}

Gravitational lensing is the effect whereby the path of light is deflected as it travels through the large-scale structure of the Universe. Lensing effects are produced primarily by the distribution of static mass in the Universe, however the movement of mass induces an additional and in general a much smaller modulation to the lensing signal. This effect is sometimes referred to as a gravitomagnetic effect \citep{schafer2006}; due to the analogy with the magnetic field within electromagnetism or alternatively the frame-dragging potential \citep{Schneider1992}.

This momentum-dependent modulation of gravitational lensing has been measured within the Solar system (\cite{Fomalont2003}, \cite{Everitt2011}). However it has not yet been observed on cosmological scales. A measurement of this subtle relativistic effect would provide a manner to directly measure the total matter momentum field (including the dark matter contribution) and could be used to test Lorentz invariance on cosmological scales.

Gravitational lensing measurements of the static mass distribution of the Universe, provides us with powerful constraints on cosmology through cosmic shear studies (for example \cite{Wright2025},\cite{Amon2022}, \cite{Dalal2023}) and mass-calibration for galaxy cluster cosmology (for example \cite{McClintock2019},\cite{Murray2022},\cite{Bellagamba2019}). However measuring the momentum field will provide us with complementary information, beyond the density field (\cite{Cai2025}).

The predicted momentum corrections to lensing are small, on the order of a factor of $v/c$ smaller than the usual lensing effects, where $v$ is the speed of moving mass and $c$ is the speed of light. Therefore measuring this effect is challenging. The observed convergence field is the sum of two contributions: the density lensing field sourced by the static mass distribution and the gravitomagnetic lensing field sourced by matter currents. By cross-correlating the total observed lensing signal with a tracer of the momentum field that is uncorrelated with the density lensing field, we can isolate the gravitomagnetic contribution. 

In this work we consider the detectability of gravitomagnetic effects by cross-correlating the lensing field with a reconstructed momentum field estimated from the galaxy overdensity field and compare this to the cross-correlation with the kinetic Sunyaev Zel'dovich (kSZ) effect field (which also traces the momentum field). We outline the observational prospects for the first cosmological-scale measurement of this effect. 

In Section \ref{sec:theory} we overview the effects of weak gravitational lensing for moving masses, the kSZ effect, and the angular spectrum of the cosmic momentum field, the cosmic lensing-momentum field and the kSZ-field. In Section \ref{sec:reconstruction} we present the reconstruction of the cosmological momentum field from the density field of galaxies as well as the reconstruction of the lensing-field from the density field of galaxies. In Section \ref{sec:simulations} we verify the validity of our predictions using ray-tracing through N-body simulations. In Section \ref{sec:detectability} we present the detectability of this signal under different observational considerations before concluding in Section \ref{sec:conclusions}.

\section{Theoretical background }
\label{sec:theory}

A useful way to conceptualise gravitational lensing is through an analogy with optical refraction. In which we can define an effective index of refraction for a gravitational lens (for example \cite{Schneider1992}). This analogy is useful provided that the Newtonian gravitational potential $\Phi$ is small and that the lensing mass distribution is slowly moving. Both of these conditions are generally met on cosmological scales given observed cluster masses (\cite{Murray2022}) and cluster velocities (\cite{Hand2012}). For a static gravitational potential, the effective index of refraction is,

\begin{equation}
    n_{\rm{eff}} = 1 - \frac{2}{c^2} \Phi.
\end{equation}

To derive the effects of a gravitational lens in motion we can consider a Lorentz transformation of the optical medium, such that we are in a reference frame with a relative motion with respect to the lens. This provides a direct analogy with the Fizeau experiment (\cite{Fizeau1851}), in which the refraction of light by a moving lens was observed. Under this transformation the effective index of refraction becomes,

\begin{equation}
\label{eq:refraction_index}
    n_{\rm{eff}} = 1 - \frac{2}{c^2} \Phi - \frac{4}{c^2} \frac{v_{\parallel}}{c} \Phi
\end{equation}

where $v_{\parallel}$ is the line-of-sight velocity of the lens. Therefore the effective index of refraction for a moving lens has a correction factor of $( 1 + 2 \frac{v_{\parallel}}{c} )$ when compared to the stationary lens. This in turn leads to a correction factor of $( 1 + 2 \frac{v_{\parallel}}{c} )$ for both the observed deflection angle from a moving lens and its convergence and shear. This equation has been derived in a more formal manner by many authors (for example \cite{Schneider1992}, \cite{schafer2006}).

\subsection{Gravitational lensing from moving masses}

From Equation~\ref{eq:refraction_index}, we can derive the weak lensing convergence in the usual manner, except with a modified source term. Instead of only the matter overdensity $\delta$, we have the addition of a term proportional to the matter current, along the line-of-sight, $j_{\parallel} \approx \delta v_{\parallel}$. Therefore, the weak lensing convergence $\kappa$ in a flat Universe for a source at a comoving distance $\chi_s$ is given by (\cite{Schneider1992}):

\begin{equation}
    \label{eq:gravitomagnetic_convergence}
    \kappa( \vec{\theta}, \chi_s ) = \frac{3 H_0^2 \Omega_m}{2c^2} \int_{0}^{\chi_s} d\chi \frac{\chi  (\chi_s - \chi )}{a(\chi ) \chi_s} \left( \delta + \frac{2}{c} j_{\parallel}\right)
\end{equation}

The convergence field can be separated into two terms, $\kappa = \kappa_{\Phi} + \kappa_{j_{\parallel}}$, where $\kappa_{\Phi}$ is the standard convergence sourced by the static mass distribution and $\kappa_{j_{\parallel}}$ is the gravitomagnetic contribution sourced by the line-of-sight momentum field. The lensing kernel as a function of comoving distance $\chi$ is,

\begin{equation}
\label{eq:lensing_kernel}
    K_{j_{\parallel}}(\chi) = \frac{3 H_0^2 \Omega_m}{c^3} \frac{\chi(\chi_s-\chi)}{a(\chi) \chi_s}.
\end{equation}

Therefore the kernel depends on the geometry of each lens-source system.

\subsection{The kinetic Sunyaev Zel'dovich effect}

The kinetic Sunyaev-Zel'dovich (kSZ) effect (\cite{Sunyaev1980}) provides an independent probe of the line-of-sight momentum field. The kSZ effect occurs when cosmic-microwave background (CMB) photons Compton scatter off of ionised electrons which have a net motion with respect to the CMB. This is due to the observed kinematic CMB dipole in the rest-frame of the ionised gas. As both the velocity of the ionised gas and the gas density are expected to follow the overall distribution of the matter field, the kSZ effect can be used to measure the projected momentum field. The kSZ effect leads to a temperature change in the CMB of, 

\begin{equation}
    \frac{\Delta T}{T_0}(\vec{\mathbf{n}}) = -\int d\chi \, e^{-\tau(\chi)} \sigma_T n_e(\vec{\mathbf{n}},\chi) \frac{v_{\parallel}}{c}
\end{equation}

where $T_0$ is the mean temperature of the CMB, $\sigma_T$ is the Thomson scattering cross-section, $\tau(\chi)$ is the Thomson optical depth to a distance $\chi$, and $n_e$ is the free electron number density. Assuming that on large scales the free electrons trace the matter distribution, we can relate the electron momentum to the matter momentum field, $n_e v_{\parallel} \approx \bar{n}_e(z) ( v_{\parallel} + j_{\parallel})$. The kSZ kernel is therefore,

\begin{equation}
    K_{\rm{kSZ}}(\chi) = - \frac{\sigma_T \bar{n}_e(\chi) e^{-\tau(\chi)}}{c}
\end{equation}

Which has a different dependence on comoving distance to the lensing kernel in Equation \ref{eq:lensing_kernel}.

\subsection{The projected cosmic momentum field}

Theoretical predictions for the power spectrum of the momentum field have been substantially studied (\cite{Ostriker1986}, \cite{Park2017}). To calculate the projected momentum field, we need the 3D power spectrum of the momentum field, $P_q(k,z)$. It can be shown that this is dominated by the rotational component of the momentum field (e.g., \cite{Ma2002}, \cite{Park2017}). \footnote{Therefore we neglect the connected four-point contribution $\langle\delta\delta v v\rangle_c$, which is subdominant over the linear scales that dominate our forecasts (roughly $\ell \lesssim 5000$ for the kernels used here) \citep{Ma2002,Barrera2022,Park2017}.}. The relevant power spectrum is therefore (\cite{Ma2002}, \cite{Barrera2022}):

\begin{equation}
\label{eq:momentum_power_spectrum}
\begin{aligned}
P_{q}(k, z) ={}& \int \frac{d^3\mathbf{k\prime}}{(2\pi)^3} (1 - \mu^2) \\
& \times \left[ P_{\delta\delta}(|\mathbf{k} - \mathbf{k\prime}|) P_{vv}(k\prime) - \frac{k\prime}{|\mathbf{k} - \mathbf{k\prime}|} P_{\delta v}(|\mathbf{k} - \mathbf{k\prime}|) P_{\delta v}(k\prime) \right],
\end{aligned}
\end{equation}

where $\mu = \hat{\mathbf{k}} \cdot \hat{\mathbf{k\prime}}$. In the linear regime, the velocity power spectrum is related to the matter power spectrum $P_{\delta \delta}(k)$ with $P_{vv}(k) = (aHf/k)^2 P_{\delta\delta}(k)$, and the velocity-density cross-power spectrum is $P_{v\delta}(k) = (aHf/k) P_{\delta\delta}(k)$.

Following \citet{hu2000} and \citet{Ma2002}, throughout the article we use the nonlinear matter power spectrum for density only terms, while using the linear theory matter power spectrum for the velocity power spectrum $P_{vv} = (aHf/k)^2 P_{\delta\delta}$ and the density-velocity cross-spectrum $P_{\delta v} = (aHf/k) P^{\rm lin}_{\delta\delta}$. This is motivated by the $1/k^2$ weighting of the velocity spectrum, which suppresses the impact of nonlinear velocity corrections relative to nonlinear density corrections \citep{Ma2002}.

For a given projection kernel, the angular cross-power spectrum of two momentum-dependent observables, A and B, is given by the Limber approximation:

\begin{equation}
\label{eq:projection}
C_{\ell}^{AB} = \int d\chi \frac{K_A(\chi) K_B(\chi)}{\chi^2} P_{AB}(k=\ell/\chi,z)
\end{equation}

where the kernel corresponding to the observable of interest can be inserted to get the desired auto or cross-correlation. The cross-correlation between two observables will be greater when the kernels are similar. 

\section{Reconstructing cosmological fields}
\label{sec:reconstruction}

In this section, we show how the galaxy density field can be used to reconstruct the momentum field and the lensing field. The interest of the former is clear, by reconstructing this field we can cross-correlate it with the lensing field to isolate the gravitomagnetic lensing field from the static mass lensing field. The interest in reconstructing the lensing field is to reduce the variance of the static lensing field on the
cross-correlation between the lensing field with the reconstructed momentum field, if this is not yet clear it should become clear in the following sections.

\subsection{ The cosmic momentum field }

The galaxy overdensity field is defined as $\delta_g = n_g / \bar{n}_g -1 $, where $n_g$ is the number of galaxies in a specified voxel and $\bar{n}_g$ is the mean number of galaxies over all of the voxels. Assuming
linear biasing of galaxies this galaxy overdensity field can be related to the underlying matter overdensity field $\delta_m$,

\begin{equation}
  \delta_g = b_g \delta_m
\label{eq:linear_bias}
\end{equation}

where $b_g$ is the linear galaxy bias (see \cite{Desjacques2018} for a review on galaxy-biasing). Within the standard model of cosmology velocity perturbations are generated purely from matter perturbations. Within
linear theory we can relate the velocity field to the matter field and in turn substitute the matter density field for the galaxy field using \ref{eq:linear_bias}, therefore in Fourier-space,

\begin{equation}
\label{eq:recon_velocity}
  \hat{\vec{v}}(\vec{k})_g = -i aHf \frac{\delta_m({k} )}{k^2} \vec{k} = -i aHf \frac{1}{b_g} \frac{\delta_g(k)}{k^2} \vec{k}
\end{equation}

where $H$ is the Hubble equation, $f$ is the linear growth function which can be written as $f=d \rm{ln} D/ d\rm{ln} a$ where $D(a)$ is the linear growth factor (\cite{Kaiser1987}).

The reconstructed momentum field in Fourier-space is,

\begin{equation}
\label{eq:recon_momentum}
  \hat{\vec{q}}(k) = \frac{iaHf}{b_g} \hat{\delta}_g({k} ) \frac{\vec{k}}{k^2} \left( 1 + \frac{\hat{\delta}_g({k} ) }{ b_g }\right)
\end{equation}

Here we have written the momentum field in terms of the galaxy density field. Therefore we can estimate the momentum field directly from the galaxy density field. There are many different methods for doing so (for
example \cite{Padmanabhan2012}, \cite{White2015}, \cite{Burden2015}). Here we consider only the simplest which is valid on linear scales, however through using non-linear information and the information held within
redshift-space distortions the velocity field reconstruction could be performed with lower noise. By then combining the density field with the reconstructed velocity field we can reconstruct the momentum field.

After obtaining the reconstructed momentum field we can project this using Equation \ref{eq:gravitomagnetic_convergence} to provide a reconstructed gravitomagnetic field for cross-correlation. Importantly we can choose
the projection kernel to match that of the observed field in order to maximise the cross-correlation of the fields. The same is not possible for the kSZ field, which is already observed in projection.

The reconstructed momentum field will not be described by the power-spectrum in Equation \ref{eq:momentum_power_spectrum} as the filtering process will change the power-spectrum (\cite{ho2009}). The reconstructed momentum field also
contains noise, therefore we need to calculate both the autocorrelation of the reconstructed momentum field and the cross-correlation of the reconstructed momentum field with the true momentum field. Both of these
power-spectra are required to estimate the signal-to-noise ratio for the detection of this cross-correlation, as explained in more detail in Section \ref{sec:detectability}. 

The Wiener filter is,

\begin{equation}
    W(k) = \frac{b_g^2 P_m(k)}{ b_g^2 P_m(k) + N(k)}
\end{equation}

where for galaxy surveys, the noise power spectrum is dominated by shot noise,

\begin{equation}
N(k) = \frac{1}{\bar{n}_g(z)}
\end{equation}

where $\bar{n}_g(z)$ is the mean number density of galaxies.

Therefore the auto-power spectrum of the reconstructed momentum field is,

\begin{equation}
\label{eq:recon_auto_momentum_power_spectrum}
\begin{aligned}
P_{\hat{q} \hat{q}}(k, z) ={}& \int \frac{d^3\mathbf{k\prime}}{(2\pi)^3} (1 - \mu^2) b_g^4 W^4(k)  \\
& \times \left[ P_{\delta\delta}(|\mathbf{k} - \mathbf{k\prime}|) P_{vv}(k\prime) - \frac{k\prime}{|\mathbf{k} - \mathbf{k\prime}|} P_{\delta v}(|\mathbf{k} - \mathbf{k\prime}|) P_{\delta v}(k\prime) \right],
\end{aligned}
\end{equation}

and the cross-power spectrum between the reconstructed momentum field and the true momentum field is,

\begin{equation}
\label{eq:recon_cross_momentum_power_spectrum}
\begin{aligned}
P_{\hat{q} q}(k, z) ={}& \int \frac{d^3\mathbf{k\prime}}{(2\pi)^3} (1 - \mu^2) b_g^2 W^2(k)  \\
& \times \left[ P_{\delta\delta}(|\mathbf{k} - \mathbf{k\prime}|) P_{vv}(k\prime) - \frac{k\prime}{|\mathbf{k} - \mathbf{k\prime}|} P_{\delta v}(|\mathbf{k} - \mathbf{k\prime}|) P_{\delta v}(k\prime) \right],
\end{aligned}
\end{equation}

Note that in the absence of noise ($N \rightarrow 0$), we recover $P_{\hat{q}q} = P_{\hat{q}\hat{q}} = P_{qq}$, where $P_{qq}$ is given by Equation \ref{eq:momentum_power_spectrum}. However, in realistic scenarios, the noise significantly suppresses the reconstructed power on small scales where $N \gg b^2_g P_{\delta \delta}(k,z)$. 

These three-dimensional power spectra can then be projected along the line of sight using the appropriate window functions to obtain the angular correlation functions needed for comparison with lensing observations. The projection integrals involve both the signal and noise contributions from the reconstructed momentum field. We can use Equation \ref{eq:projection} to calculate the relevant projected angular correlation functions for the noise fields.

\subsection{ The cosmic convergence field }
\label{subsec:cosmic_convergence_field}

In addition to reconstructing the momentum field we also have an interest in reconstructing the cosmic convergence field. As we will see in Section \ref{sec:detectability} the primary source of noise for future lensing
experiments on the gravitomagnetic lensing term will be the cosmic variance from the density convergence field.

In principle the convergence field can also be estimated from the galaxy overdensity field. We take equation \ref{eq:gravitomagnetic_convergence} and ignoring the gravitomagnetic contribution, as we wish only to reconstruct here the density convergence we have,
\begin{equation}
\label{eq:convergence_reconstruction}
  \hat{\kappa}_{\Phi}( \vec{\theta} , \chi ) = \frac{3 H_0^2 \Omega_m}{2c^2} \int_{0}^{\chi} d\chi^{\prime} \frac{\chi^{\prime}  (\chi - \chi^{\prime} )}{a(\chi^{\prime} ) \chi} \hat{\delta}_g \frac{1}{b_g}
\end{equation}

where once again we simply replace the matter density field with the galaxy density field using the linear galaxy bias relation $\delta_g=b_g\delta_m$. This can be then used to subtract the density lensing term from the
observed lensing signal. For example $\kappa -\hat{\kappa}_{\Phi}$ will have a smaller density lensing convergence as compared to the directly observed field $\kappa$. However we will see that in practice, even in an
idealised case where the galaxy bias is perfectly known that this reconstruction is of limited use.

\section{ Simulations }
\label{sec:simulations}

The theoretical predictions in the previous section are valid within the linear approximation. However, real cosmological fields are non-linear, in particular on small scales. These non-linearities will impact the efficiency of the momentum field reconstruction and the details of the cross-correlation between the different cosmological fields. To accurately assess these effects and verify that our reconstructed momentum field remains negligibly correlated with the density lensing field even on non-linear scales – a critical requirement for isolating the gravitomagnetic signal – we use N-body simulations. This provides a test that the correlation is insignificant even on non-linear scales. Including all the possible contributions from higher-order correlations.

In this work we use the Quijote simulations \footnote{Accessed through https://quijote-simulations.readthedocs.io/en/latest/index.html}. These are dark matter-only N-body simulations, and we therefore ignore the important intricacies of the galaxy-halo connection, the halo-matter connection and baryonic effects. The Quijote simulations provide 512$^3$ particles within a 1000 Mpc/h box, allowing for the sampling of a wide range of scales. We use five different redshift snapshots: z=0.0,0.5,1.0,2.0, and 3.0.

\subsection{ Constructing the fields }

We use a simplified snapshot geometry (\cite{Smith2018}) instead of creating realistic light-cones (\cite{Breton2019}, \cite{Cai2017}, \cite{Zhu2017}). Therefore our setup does not include important light-cone effects, we can nonetheless use these fields to confirm the accuracy of our theoretical power spectra and angular correlation functions.

From the list of particle positions and velocities provided by the simulation snapshots, we construct 3D grids of matter overdensity, $\delta$, and momentum, $\vec{q}$, using a cloud-in-cell mass assignment scheme (implemented within \cite{Pylians} \footnote{https://pylians3.readthedocs.io}). From these 3D grids we calculate 2D fields by projecting along an axis of the box, using the relevant kernel.

For the density convergence field, $\kappa_{\Phi}$, 

\begin{equation}
\label{eq:convergence_projection_scalar}
\kappa( \theta ) = \frac{3}{2} \frac{H_0^2}{c^2} \Omega_{m,0} \sum_i \frac{x_i (x_s-x_i)}{x_s} \delta(x_i)
\end{equation}

\begin{figure*}
\centering
   \includegraphics[width=18cm]{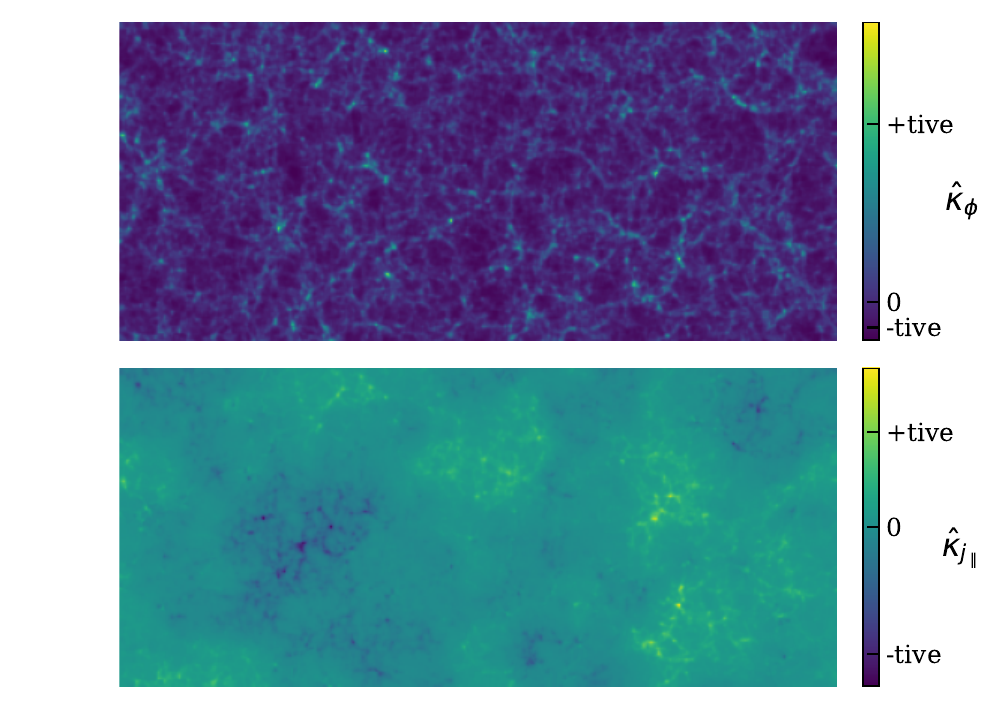}
     \caption{ Projected convergence fields constructed from the Quijote simulations using a snapshot geometry. Top: The density} convergence field, $\kappa_{\Phi}$. Bottom: the momentum convergence field, $\kappa_{\parallel}$. Both images have been smoothed with a Gaussian filter. 
    \label{fig:simulated_fields}
\end{figure*}

For the gravitomagnetic convergence field, $\kappa_{j_{\parallel}}$,

\begin{equation}
\label{eq:convergence_projection_momentum}
\kappa_{j_{\parallel}}( \theta ) = \frac{3 H_0^2}{c^3} \Omega_{m,0} \sum_i \frac{x_i (x_s-x_i)}{x_s} j_{\parallel}(x_i)  .
\end{equation}

where the sum runs along the specified line-of-sight. Subregions of fields created in this way are shown in Figure \ref{fig:simulated_fields}. There are a couple of interesting points to be noted in this Figure. There are overdensities, clear large positive regions in the density convergence field where the momentum convergence field is negative. This is because there are parts of the cosmic web which are moving towards or away from the observer. An overdensity moving away from the observer will appear negative in the momentum convergence field. Conversely there are underdense (negative $\kappa_{\Phi}$) regions in the density convergence which appear positive in the momentum convergence field, $\kappa_{\parallel}$, these are underdense regions moving away from the observer.

\subsection{ Reconstructed momentum field }

We can also reconstruct the momentum field directly from the density field, without using the particle velocity information, in order to mimic the effects of reconstruction where only the galaxy density information is available. This allows us to verify that the reconstruction method does not introduce a correlation between the density convergence field and the reconstructed momentum convergence field, which would complicate the detection of the momentum convergence signal.

The reconstruction procedure is performed as follows,

\begin{itemize}
    \item We start with the 3D matter over-density field $\delta(\vec{x})$ computed from the simulation snapshot
    \item We compute its Fourier transform to obtain $\delta(\vec{k})$
    \item We compute the linear-theory velocity field in Fourier space using equation \ref{eq:recon_velocity}, assuming $b_g=1$
    \item We compute the inverse Fourier transform to obtain $\hat{v}(\vec{x})$
    \item Finally we multiply this by the matter density field $1+\delta(\vec{}x)$ in order to obtain the reconstructed momentum field $\hat{j}(\vec{x})$
\end{itemize}

As before, this 3D reconstructed momentum field is then projected using the gravitomagnetic lensing kernel (Equation \ref{eq:convergence_projection_momentum}), to provide the reconstructed momentum lensing field $\hat{\kappa}_{j_{\parallel}}$.

Therefore we have two fields the momentum convergence which are essentially $\kappa_{j_{\parallel}} \sim \sum  \delta v_{\parallel}$ and the reconstructed momentum convergence $\hat{\kappa}_{j_{\parallel}} \sim \sum  \hat{\delta} \hat{v}_{\parallel}$, which is composed of the reconstructed density $\hat{\delta}$ (here actually exact, but in practice this would be estimated from the galaxy density field) and the reconstructed velocity $\hat{v}$.

\begin{figure*}
\centering
   \includegraphics[width=18cm]{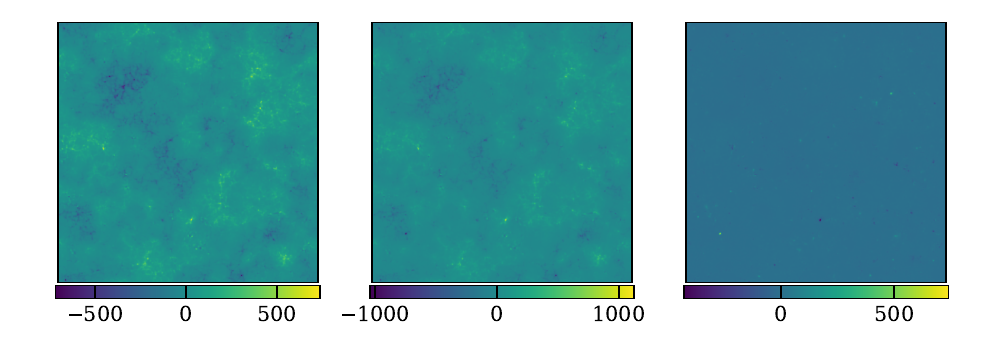}
     \caption{ Left: the simulated gravitomagnetic field. Middle: the gravitomagnetic field reconstructed from the density field. Right: the difference between the gravitomagnetic field and the gravitomagnetic field reconstructed from the density field. Both fields have been smoothed using a Gaussian filter. }
    \label{fig:recon_difference_images}
\end{figure*}

In Figure \ref{fig:recon_difference_images} we see an excellent agreement between the two fields; the convergence momentum field and the convergence momentum field that has been reconstructed from the density field. In particular the large-scale features are well reconstructed however the small scale features are different. This is precisely as we would expect, whereby the large-scales are well described by the linear model in Equation \ref{eq:recon_momentum} and the small scales become non-linear therefore the approximation of linearity is broken.

To quantify the quality of the reconstruction and the lack of correlation between the reconstructed momentum convergence field and the density convergence field, we compute the cross-correlation coefficient $r(k) = C_{AB}(k) / \sqrt{C_{AA}(k) C_{BB}(k)}$ for different pairs of projected fields in Fourier space. We perform this analysis for 15 different realisations by using 5 redshift snapshots and projecting along the 3 axes of the snapshots for each snapshot.

\begin{figure}
	\includegraphics[width=\columnwidth]{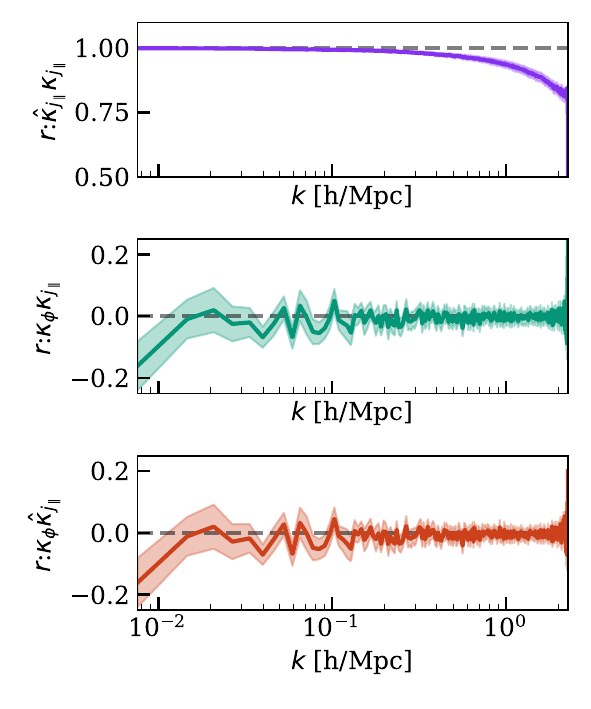}
    \caption{ The correlation coefficient $r(k)$ between each of the projected fields.  These tests are diagnostics to show that higher-order correlations that we do not model do not introduce significant correlations between the density lensing and matter current lensing fields. The solid lines show the mean correlation coefficient from 15 different realisation of the fields (5 snapshots $\times$ 3 projection axes), the shaded regions show $1\sigma$ error bars calculated from the dispersion of the estimated correlation coefficients. Top: the correlation between the reconstructed momentum convergence $\hat{\kappa}_{j_{\parallel}}$ and the momentum convergence $\kappa_{j_{\parallel}}$. Middle: the correlation between the momentum convergence $\kappa_{j_{\parallel}}$ and the density convergence $\kappa_{\phi}$. Bottom: the correlation between the density convergence $\kappa_{\phi}$ and the reconstructed momentum convergence $\hat{\kappa}_{j_{\parallel}}$.}
    \label{fig:r_correlations}
\end{figure}

We calculate this for the cross-correlation between the reconstructed momentum convergence field $\hat{\kappa}_{j_{\parallel}}$, the momentum convergence field $\kappa_{j_{\parallel}}$ and the density convergence field $\kappa_{\phi}$. The results of these cross-correlations are shown in Figure \ref{fig:r_correlations}. We observe an excellent correlation between the reconstructed momentum convergence field and the true momentum convergence field on large scales. Demonstrating that in this simplified scenario it is possible to reconstruct accurately the gravitomagnetic lensing field from the 3D density field. The correlation rapidly decreases at smaller scales, consistent with the breakdown of the linear theory approximation in our reconstruction method. Importantly we measure no correlation between the reconstructed momentum convergence field with the density convergence field. This is a crucial result, confirming that our linear reconstruction method does not introduce a spurious correlation with the dominant density lensing signal, thus allowing for a clean isolation of the gravitomagnetic effect. The robustness of this null correlation is a critical requirement for detecting the tiny gravitomagnetic lensing signal, as any leakage from the dominant density convergence could easily overwhelm it.

\section{ Detection forecasts }
\label{sec:detectability}

Having validated our theoretical framework and confirmed the negligible cross-correlation between the reconstructed momentum field and the density lensing field through simulations, we now proceed to forecast the observational prospects for detecting gravitational lensing induced by cosmological matter currents. This section details the signal-to-noise ratio estimations for various observational configurations.

\subsection{Signal-to-Noise Estimations}

The signal-to-noise ratio (S/N) for the cross-power spectrum of two Gaussian fields is,

\begin{equation}
    \left(\frac{S}{N} \right)^2 = f_{\rm{sky}} \sum_{\ell} (2 \ell+1) \Delta \ell \frac{\left(C^{ AB} \right)^2}{ C^{A} C^B + (C^{AB})^2 } .
        \label{eq:snr}
\end{equation}

where $A$ and $B$ are to be replaced by the field of interest (for example $\kappa$, $\kappa_{j_{\parallel}}$, $\hat{\kappa}_{j_{\parallel}}$). The power spectra here are the observed power spectra including the noise and in the following we assume that there is no correlation between the different noise fields and $f_{\rm{sky}}$ is the sky fraction. The power spectra $C$ in Equation \ref{eq:snr} are obtained from the 3D power-spectra using the Limber equation, as detailed in Equation \ref{eq:projection}.

\subsection{ Convergence-reconstructed momentum correlations }

Firstly we consider the cross-correlation of the lensing convergence with the reconstructed momentum convergence. The results are shown in the left panel of Figure \ref{fig:snr_forecasts}. We consider a range of galaxy densities for the reconstruction of the momentum lensing field. From $\bar{n}_g=10^{-4}$ Mpc$^{-3}$, which achievable to high redshifts from current surveys to $\bar{n}_g=10^{-2}$ Mpc$^{-3}$, which is quite an ambitious value for future spectroscopic galaxy surveys \footnote{Although the DESI bright galaxy survey sample has already achieved such densities at relatively low redshift. \citep{desi2025_sample}}. For the noise on the lensing signal we consider both Stage-4 lensing survey noise (\cite{euclid1}, \cite{2009arXiv0912.0201L}), which corresponds to a galaxy density of 30 per square arcminute, and cosmic variance noise. In each case the sky-fraction is set to $f_{\rm{sky}} \approx 0.25$ and the source redshift for lensing at $z=2$.

In each of the following we consider the $S/N$ ratio for a maximum analysis multipole of $\ell=5000$ and the Stage-4 lensing noise. At such small scales we do not expect our theoretical power-spectra to be precise, but here we are interested in the order of magnitude of the effect as opposed to a precise prediction.

The results are as follows:

\begin{itemize}
    \item For $\bar{n}_g=10^{-4}$ Mpc$^{-3}$ we forecast an $S/N \sim 2.0$, therefore the effect would not be detectable.
    \item For $\bar{n}_g=10^{-3}$ Mpc$^{-3}$ the  $S/N \sim 5.7$, which indicates that a statistically significant detection is possible.
    \item For the ambitious galaxy density of $\bar{n}_g=10^{-2}$ Mpc$^{-3}$ we achieve an $S/N \sim 9.5$. This would represent a robust, statistically significant detection of the gravitomagnetic lensing signal. 
\end{itemize}

The results for cosmic variance limited noise and Stage-4 lensing noise are similar in our analysis. This is primarily because our current analysis does not consider multiple redshift bins, therefore the noise on the lensing is principally from the cosmic variance of the density lensing, rather than shape-noise limited on most angular scales.

\subsection{ Cosmic variance limits }

In order to understand the limitations of the detection of the gravitomagnetic lensing field we consider the cosmic variance limits for different types of cross-correlations involving tracers of the projected momentum field. These results are shown in the middle panel of Figure \ref{fig:snr_forecasts}.

We consider three tracers of the projected momentum field: a noiseless gravitomagnetic field, the reconstructed momentum field (with $\bar{n}_g=10^{-2}$ Mpc$^{-3}$) and the kSZ field. The use of the kSZ field for measuring the gravito-magnetic signal has been considered in detail by \cite{Barrera2022}. We find similar results to this work. The kSZ signal, in the presence of the primary CMB, contains signal on small scales. Relatively with the reconstructed momentum field from the galaxy density we are able to recover the signal on large-scales. We see that the reconstructed momentum field approaches a similar SNR to the noiseless momentum field. This is because the noise is primarily from the cosmic variance contribution from the dominant density lensing field. Therefore we should consider how to reduce the impact of the density lensing field.

\subsection{ Density convergence subtracted - reconstructed momentum correlations }

The detection of this signal is limited by the gravitomagnetic convergence being considerably smaller than the density convergence field. Therefore ideally we would remove the density convergence field from the observed convergence. In principle this can be achieved by reconstructing the density convergence field, $\hat{\kappa}_{\phi}$, and then subtracting this from the observed convergence field $\kappa$. 

If this reconstruction was perfect the resultant field would contain only the momentum convergence contribution. We could then correlate the two fields $\kappa - \hat{\kappa}_{\phi}$ and $\hat{\kappa}_{j){\parallel}}$. This will reduce the cosmic variance contribution from the density convergence to the cross-correlation between the reconstructed momentum and the observed convergence, however it will add shot noise from the reconstruction of $\hat{\kappa}_{\phi}$. As described in Subsection \ref{subsec:cosmic_convergence_field} we a assume a simple linear bias and perfect knowledge of the galaxy bias, both of which are unreasonable assumptions. 

\begin{figure*}
\centering
   \includegraphics[]{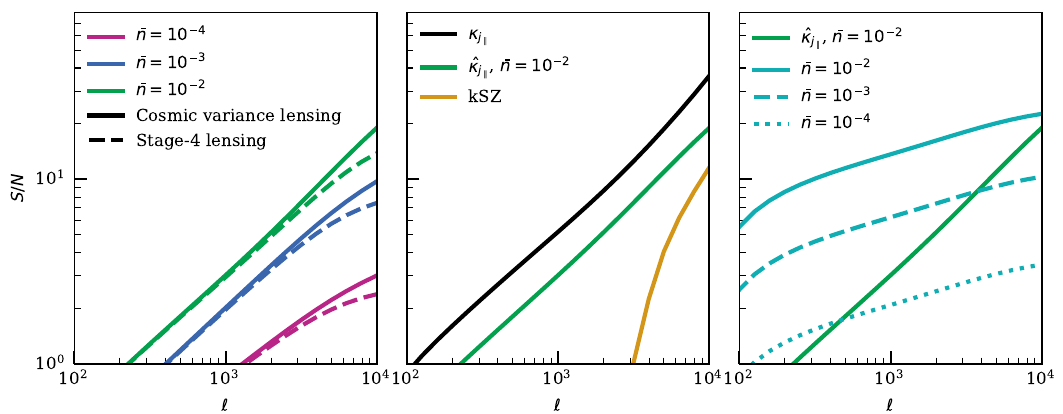}
     \caption{ Cumulative signal-to-noise ratio (S/N) forecasts for detecting the gravitomagnetic lensing signal as a function of maximum angular multipole $\ell$. All forecasts assume a sky fraction of $f_{\rm{sky}} \approx 0.25$. Left: the S/N for the cross-correlation of the lensing convergence with the reconstructed momentum convergence field for different galaxy number densities ($\bar{n}_g$, in units of Mpc$^{-3}$); solid lines show cosmic variance limited lensing noise, while dashed lines show Stage-4 lensing survey noise. Middle: the cosmic variance limited S/N for three different momentum field tracers: the true, noiseless field ($\kappa_{j){\parallel}}$), the reconstructed field ($\hat{\kappa}_{j){\parallel}}$) assuming a galaxy density of $\bar{n}_g = 10^-2$ Mpc$^-3$, and the kinetic Sunyaev-Zel'dovich (kSZ) effect. Right: the S/N for the cross-correlation after subtracting a reconstructed density convergence field ($\hat{\kappa}_{\phi}$), where the different lines show the effect of varying the galaxy density used for the density reconstruction.  }
    \label{fig:snr_forecasts}
\end{figure*}

The results for different galaxy densities in the density convergence reconstruction are shown in the right panel of Figure \ref{fig:snr_forecasts}. We can see that on large-scales where the noise is dominated by the cosmic variance from the density lensing we improve upon on our SNR. On small scales the SNR is decreased as the shot-noise from the convergence reconstruction adds more noise to the cross-correlation. Therefore whilst the improvements are modest at large multipoles the approach can significantly improve our SNR on linear scales where modelling is easiest.

\section{Conclusions}
\label{sec:conclusions}

We have investigated the prospects for the detection of gravitational lensing induced by cosmological matter currents, a subtle relativistic effect known as gravitomagnetic lensing. Our results show that while measuring this signal is undoubtedly challenging, a statistically significant detection is within reach of upcoming and future cosmological surveys.

The key to isolating this effect, which is typically orders of magnitude smaller than the standard lensing signal from static mass, lies in cross-correlating the total lensing convergence field with a tracer of the large-scale momentum field. We demonstrated through N-body simulations that the gravitomagnetic field can be accurately reconstructed from the density field observed in large-scale structure surveys. Crucially, our analysis confirms that this reconstructed momentum field is uncorrelated with the dominant density lensing signal, a vital condition for a clean measurement. This allows the cross-correlation to effectively filter out the much larger density lensing contribution, isolating the gravitomagnetic signal.

Our forecasts show that a Stage-4 CMB lensing experiment combined with a spectroscopic galaxy survey with a number density of $\bar{n}_g=10^{-3}$ Mpc$^{-3}$ can achieve a detection with a signal-to-noise ratio (S/N) of approximately 5.7. More ambitious surveys could yield an even more robust detection with an S/N approaching 10. We also found that the primary limitation for this measurement is the cosmic variance from the density lensing field itself. This suggests that techniques to mitigate this variance, such as subtracting a reconstructed density lensing map, could offer modest but valuable improvements, particularly on large, linear scales where cosmological models are most reliable.

A future detection would represent a novel test of general relativity on cosmological scales and provide a new, independent way to probe the cosmic velocity field, including the motion of dark matter. While our analysis relies on angular power spectra, the S/N could be further enhanced by employing more advanced statistical methods and by leveraging tomographic information from multiple redshift bins. Such a measurement would open a new window onto the dynamics of the universe and the fundamental nature of gravity

\begin{acknowledgements}{Calum Murray thanks Cotonou Alvarez Cardona for a careful reading of the manuscript and providing valuable comments. We thank the anonymous referee for helpful comments and questions that greatly improved the clarity of the text. }
\end{acknowledgements}

\bibliographystyle{aa}
\bibliography{example} 

\end{document}